\documentclass[aps,pra,showpacs,floatfix,twocolumn]{revtex4}
\usepackage{graphicx}
\usepackage{dcolumn}
\begin{document}
\title{State-insensitive bichromatic optical trapping}
\author{Bindiya Arora}\altaffiliation[Permanent address: ]
 {Department of Physics, Guru Nanak Dev University, Amritsar 143005, Punjab, India}
\affiliation{Department of Physics and Astronomy,
 University of Delaware, Newark, Delaware 19716}
\author{M. S. Safronova}
\affiliation{Department of Physics and Astronomy,
University of Delaware, Newark, Delaware 19716}
\author{Charles W. Clark}
 \affiliation{
 Joint Quantum Institute, National Institute of Standards and Technology
and the University of Maryland, Gaithersburg, Maryland 20899-8410, USA}
\date{\today}
\begin{abstract}
We propose a scheme for state-insensitive trapping of neutral atoms
by using light with  two independent wavelengths. In particular, we
describe the use of trapping and control lasers to minimize the
variance of the potential experienced by a trapped Rb atom in ground
and excited states. We present calculated values of wavelength pairs for which the $5s$ and
$5p_{3/2}$ levels have the same ac Stark shifts in the presence of two laser fields.
\end{abstract}
 \maketitle
\section{Introduction}

The emerging advantages of using transitions between different atomic electronic configurations for frequency standards
and quantum information processing
 are accompanied by disadvantages with respect to previous methods in which
 atomic qubits were hyperfine states of the same configuration.
  In the latter case, the energy shifts of states induced by external trapping
 fields are small, and can be calculated to an accuracy that usually does
not make a significant contribution to  the uncertainty budget. When qubits are associated with different
configurations, on the other hand, the magnitude and even the sign of differential field shifts are uncontrolled in the
first instance.

The ability to trap neutral atoms inside high-Q cavities in the strong
 coupling regime is of particular importance for quantum computation and
 quantum communication schemes, where
  it is essential to precisely localize and control neutral
atoms with minimum decoherence.  In a far-detuned
 optical dipole trap, the potential experienced by an atom in its ground state can be either attractive
 or repulsive, with respect to the location of peak light intensity, depending on the sign of the ac Stark shift due to the trapping light.
  Excited-state atoms in the same trap may experience an ac Stark shift with an opposite sign,
   which affects the fidelity of experiments in which excited states are temporarily occupied,
   such as the implementation of the Rydberg quantum gate \cite{ref-n4,ref-n5,ref-n1,ref-n2,ref-n3}.

The same problem, i.e. different Stark shifts of two states,
    affects optical frequency standards based on atoms
   trapped in optical lattices, because it can introduce a significant dependence of the measured frequency of the clock transition upon the lattice wavelength.
   Katori \textit{et al.}~\cite{katori} proposed the idea of using a trapping laser tuned to a "magic wavelength'', $\lambda_{\rm{magic}}$,
at which the ac Stark shift of the clock transition is eliminated. The magic wavelength of the  $^{87}$Sr
$^1S_0-~^3P^{\circ}_0$ clock transition
  was found to be 813.5$\pm$0.9~nm in Ref.~\cite{katori1} by investigating the wavelength dependence of the carrier
  linewidth.
  This magic wavelength was later determined with even higher precision to be 813.42735(40)~nm~\cite{Sr-new}.
In a cavity quantum electrodynamics experiment, McKeever \textit{et~al.}~\cite{McKeever} demonstrated state-insensitive
trapping of Cs atoms at $\lambda_{\rm{magic}}$  $\approx$ 935 nm while still maintaining a strong coupling for the
$6p_{3/2}-6s_{1/2}$ transition.

Magic wavelengths for $np-ns$ transitions in alkali-metal atoms
from Na to Cs have been previously calculated by Arora \emph{et
al.}~\cite{magic}, using a relativistic all-order method. This was
accomplished by matching the ac polarizabilities of the atomic
$np_j$ and $ns$ states. The data in Ref.~\cite{magic} provide a wide
range of magic wavelengths for alkali-metal atoms. In the case of
the $np_{3/2}-ns$ transitions, the magic wavelengths need to be
determined separately for the $m_j=\pm 1/2$ and $m_j=\pm
3/2$ states, due to the rank-2 tensor contribution to
the polarizability of the $np_{3/2}$ level. Furthermore, there
is a substantial reduction in the number of magic wavelengths for
the $m_j=\pm 3/2$ states due to selection rules for linear
polarization. For instance, three out of the six values of
$\lambda_{\rm{magic}}$ suggested for the $5p_{3/2}-5s$ transition in
Rb are present only for the $m_{j}=\pm 1/2$ states. In such
cases, the magic wavelength becomes dependent on the particular
hyperfine state of the atom. Some of the magic wavelengths are also
in regions that are inconvenient for  present laser technology.
 Out of the remaining three
wavelengths considered in \cite{magic}, the $\lambda_{\rm{magic}}$ at 791~nm has opposite signs for the Stark shifts
for $m_j=\pm3/2$ and $m_j=\pm1/2$ states, which makes this wavelength of limited practical use. The second magic
wavelength at 776~nm is in close proximity to the Rb $5p-5d$ resonance transition at 775.8~nm, which could
mediate undesired two-photon transitions.  The third magic wavelength at 637 nm exists for all states, but its
corresponding polarizability is too small for convenient trapping (it is -500~$a^3_0$, where $a_0$ is the Bohr radius).
In summary, the single-laser scheme offers few cases in which the magic wavelengths are convenient for
state-insensitive trapping of Rb atoms~\cite{magic}.

In this paper, we investigate an obvious mechanism for remediating uncontrolled frequency shifts in transitions between
different configurations: the application of a second, "control," optical field to a system of optically-trapped atoms.
 We outline the general principles of this approach, and apply it in detail to some cases in Rb.
  Rubidium is chosen because it offers a baseline of comparison with previous, monochromatic,
attempts at control, and this serves to illustrate advantages of the bichromatic approach
 that we believe will have wide applicability.

Specifically, we find the combinations of two  wavelengths that allow to match ac Stark shifts of the Rb atom in $5s$
 and $5p_{3/2}$ states.
   In this scheme, a combination of trapping and control lasers
    allows one to minimize the difference in the trapping potentials experienced by
    the atom in ground and excited states.
This approach significantly increases the number of wavelengths at which state-insensitive trapping experiments can be
conducted.

The first step in the realization of this scheme is to calculate the
Stark shifts of the $5s$ and $5p_{3/2}$ states of the Rb atom as a
function of frequency. We use the relativistic all-order method
~\cite{relsd,magic,paper1} for the calculation of reduced
electric-dipole matrix elements involved in the evaluation of
frequency-dependent polarizabilities. In the second step, we
calculate the shift in energies of atomic states as a function of
the two laser frequencies. The wavelengths are determined where the
ac Stark shifts of the $5s$ and $5p_{3/2}$ levels
 match according to criteria that are described below.
  Several specific cases are illustrated in detail.

 \begin{table*}
\caption{\label{tab1}Magic combinations of the trap and control wavelengths $\lambda_{\rm{1}}$,
$\lambda_{\rm{2}}=2\lambda_{\rm{1}} $ for the $5p_{3/2} m_j -5s$ transition in Rb and the corresponding sum of
polarizabilities at these wavelengths. The wavelengths (in vacuum) are given in nm and polarizabilities are given in
atomic units. ${\epsilon}_1^2$ and ${\epsilon}_2^2$ represent the intensities of the two laser beams, respectively.
$\alpha^{\rm{sum}}=\alpha_1+(\epsilon_2/\epsilon_1)^2 \alpha_2$, so that the energy level shift is proportional to
$\alpha^{\rm{sum}} \epsilon_1^2$.}
\begin{ruledtabular}
\begin{tabular}{ccllll}
\multicolumn{1}{c}{$(\epsilon_2/\epsilon_1)^2$} & \multicolumn{1}{c}{$|m_j|$}&
 \multicolumn{1}{l}{$\lambda_{\rm{1}}$} &
\multicolumn{1}{l}{$\lambda_{\rm{2}}$} & \multicolumn{1}{l}{$\alpha^{\rm{sum}}(5s)$}&
\multicolumn{1}{l}{$\alpha^{\rm{sum}}(5p_{3/2})$}\\
\hline
1   &   1/2 &788        &1576       & 4990(18)      & 4914(190)\\
1   &   3/2 & 785       & 1570      & 13240(26)         & 13332(214)\\[0.5pc]
1   &   1/2 &814    &1628           & 5189(5)       & 5194(94)\\
1   &   3/2 &810        &1620        & 6086(6)       & 6070(100)\\[0.5pc]
2   &   1/2 &784.3  &1568.6          &16819(30)         &16069(436)\\
2   &   3/2 &782.7  &1565.4 &30086(50)          &29715(470)\\[0.5pc]
2   &   1/2 &798.5  &1597       &17189(18)      &17297(260)\\
2   &   3/2 &799        &1598   &15611(16)          &15874(260)\\[0.5pc]
3   &   1/2 & 782.9     &1565.8     &28017(46)          &27317(700)\\
3   &   3/2 &781.9  &1563.8     &46328(70)      & 46387(740)\\[0.5pc]
3   &   1/2 &796.8  &1593.6 &29026(34)         &29336(410)\\
3   &   3/2 &797.2  &1594.4     &24820(28)          &25059(402)\\[0.5pc]
1    &  1/2  &715    &1430       & -1047(2)          &-1031(84)\\
1    &  1/2  & 974-978       & 1948-1956   &  1271(1) - 1257(1) &  1279-1240(34)\\
1/2 &   1/2  &727        &1454       &-1641(2)           &-1648(55) \\
1/2 &   3/2  &787.4      &1574.8         &6109(20)           &6054(100)\\
1/3 &   1/2  &736        &1472   &-2113(3)           &-2094(50)\\
1/3 &   1/2  &748        &1496             & -2963(4)                  & -2988(80) \\
1/3 &   3/2  &576        &1152       &-152(1)        &-153(34)\\
1/3 &   3/2  &639        &1278       &-425(1)            &-422(44)\\
\end{tabular}
\end{ruledtabular}
\end{table*}
\section{Frequency-dependent polarizability}~\label{sec-pol}
The second-order energy shift $\Delta E$ of a monovalent atom in a state $v$ is parameterized as the sum of scalar
$\alpha_0(\omega)$ and tensor $\alpha_2(\omega)$ polarizabilities
    \begin{equation}
    \Delta
  E=-\frac{1}{2}\alpha_0(\omega)\epsilon^2-\frac{1}{2}\alpha_2(\omega)\frac{3m_j^2-j_v(j_v+1)}{j_v(2j_v-1)}~\epsilon^2,
    \end{equation}
\noindent where the laser frequency $\omega$  is assumed to be several linewidths off-resonance, $j_v$ is the angular
momentum, $\epsilon$ is the rms magnitude of the electric field, and the polarization vector of the linearly polarized
light defines the $z$ direction. The valence contribution to frequency-dependent scalar and tensor polarizability is
evaluated as the sum over intermediate $k$ states allowed by the electric-dipole transition rules~\cite{1}
\begin{eqnarray}
    \alpha_{0}^v(\omega)&=&\frac{2}{3(2j_v+1)}\sum_k\frac{{\left\langle k\left\|d\right\|v\right\rangle}^2(E_k-E_v)}{     (E_k-E_v)^2-\omega^2}, \label{eq-1} \nonumber \\
    \alpha_{2}^v(\omega)&=&-4C\sum_k(-1)^{j_v+j_k+1}
            \left\{
                    \begin{array}{ccc}
                    j_v & 1 & j_k \\
                    1 & j_v & 2 \\
                    \end{array}
            \right\} \nonumber \\
      & &\times \frac{{\left\langle
            k\left\|d\right\|v\right\rangle}^2(E_k-E_v)}{
            (E_k-E_v)^2-\omega^2} \label{eq-pol},
\end{eqnarray}
             where $C$ is given by
\begin{equation}
            C =
                \left(\frac{5j_v(2j_v-1)}{6(j_v+1)(2j_v+1)(2j_v+3)}\right)^{1/2} \nonumber
\end{equation}
and ${\left\langle k\left\|d\right\|v\right\rangle}$ are the reduced electric-dipole matrix elements. The experimental
energies $E_i$ of the dominant states contributing to this sum have been compiled for the alkali atoms in
Refs.~\cite{NIST, NIST2}.
 In addition to the scalar and tensor valence contributions,
there is a scalar core contribution  to the polarizability, $\alpha_{\rm{core}}$. For the frequency range considered in
this work, $\alpha_{\rm{core}}$ has a very small $\omega$ dependence. The static core polarizability value calculated
using a random-phase approximation \cite{datatab2} has been used in our calculations without loss of accuracy, i.e.
uncertainty of this term gives negligible contribution to the total uncertainty.

Unless stated otherwise, we use atomic units (a.u.) for all matrix elements and polarizabilities throughout this paper:
the numerical values of the elementary charge, $e$, the reduced Planck constant, $\hbar = h/2 \pi$, and the electron
mass, $m_e$, are set equal to 1. The atomic unit for polarizability can be converted to SI units via
$\alpha/h$~[Hz/(V/m)$^2$]=2.48832$\times10^{-8}\alpha$~(a.u.), where the conversion coefficient is $4\pi \epsilon_0
a^3_0/h$ and the Planck constant $h$ is factored out in order to provide direct conversion into frequency units; $a_0$
is the Bohr radius and $\epsilon_0$ is the electric constant.

    \begin{figure}[htb!]
  \includegraphics[width=3.6in]{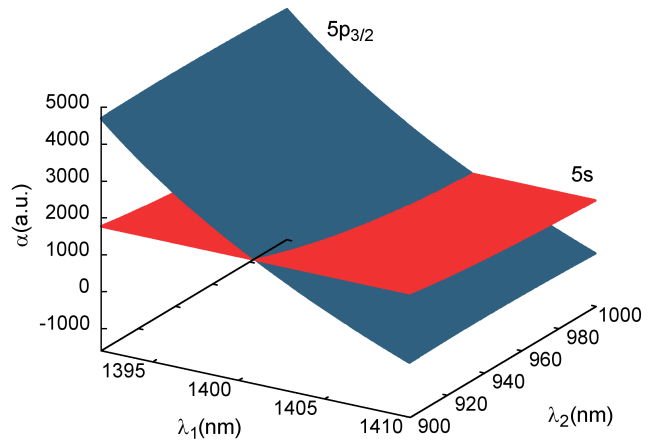}
  \caption{Surface plot for the $5s$ and
$5p_{3/2}\mbox{ }m_j = \pm 1/2$ state polarizabilities  as a function of laser wavelengths $\lambda_1$ and $\lambda_2$
for equal intensities of both lasers.}
  \label{figa}
\end{figure}

\begin{figure}[htb!]
  \includegraphics[width=3.6in,angle=0]{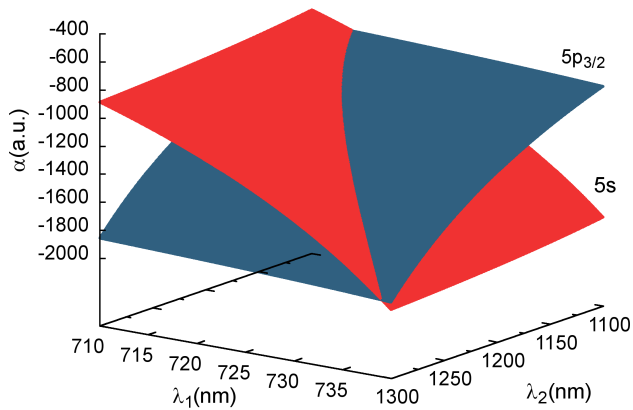}
  \caption{Surface plot for the $5s$ and
$5p_{3/2}\mbox{ }m_j = \pm 3/2$ state polarizabilities as a function laser wavelengths $\lambda_1$ and $\lambda_2$ for
equal intensities of both lasers.}
  \label{figb}
\end{figure}

The ground and excited state ac polarizabilities of the alkali-metal
atoms were previously calculated accurately in
Refs.~\cite{magic,paper1,pol-andrei,poldere}. Detailed description
of the polarizability calculations for atomic Rb is given in
Refs.~\cite{paper1,magic}. Briefly, the sums over intermediate
states $k$ in the formulas are separated
  into a dominant part $\alpha_{\rm{main}}$ that contains the first few terms
 and a remainder
$\alpha_{\rm{tail}}$. In our Rb calculations, we include all $ns$ states up to $10s$ and all $nd$ states up to $9d$ in
the $\alpha_{\rm{main}}$ term. The $\alpha_{\rm{tail}}$ contribution is calculated in the Dirac-Fock (DF)
approximation. We use a complete set of DF wave functions on a nonlinear grid generated using B-splines~\cite{relsdrb}
constrained to a spherical cavity.  A cavity radius of 220~$a_0$ is chosen to accommodate all valence orbitals of
$\alpha_{\rm{main}}$. The basis set consists of 70 splines of order 11 for each value of the relativistic angular
quantum number $\kappa$.

In the calculation of the main term, the $5p_{3/2} - 5s$ matrix elements are taken from Ref.~\cite{volz}, and the
$5p_{3/2}-4d_{j}$ E1 matrix elements are taken to be the recommended values derived in Ref.~\cite{rbk} from the Stark
shift measurements reported in ~\cite{stk}. We use the all-order method (linearized version of the coupled cluster
approach), which sums infinite sets of many-body perturbation theory terms, for the calculation of all other matrix
elements in the dominant part, $\alpha_{\rm{main}}$. Detailed description of the all-order method is given in
Refs.~\cite{CC,relsd}. For some matrix elements, it was possible to carry out semi-empirical scaling of the all-order
values to include some additional important higher-order corrections. The scaling procedure has been described in
Refs.~\cite{relsd,CC2,1}. The resulting frequency-dependent polarizabilities are used to find convenient combinations
of trap and control laser wavelengths that yield the same ac Stark shift for Rb atoms in the ground and excited
$5p_{3/2}$ levels.

\section{Results}~\label{sec-res}

In this section, we list a few appropriate combinations found for
control and trap laser wavelengths where the $5s$ and $5p_{3/2}$
state polarizabilities of Rb are closely matched. For monochromatic
light, a magic wavelength is represented by the point at which two
curves, $\alpha_{5s}(\omega)$ and $\alpha_{5p}(\omega)$, intersect
as a function of the frequency, $\omega$. In the bichromatic case,
on the other hand, we have two additional degrees of freedom, the
control frequency and the ratio of laser intensities. Thus,
bichromatic magic wavelengths are represented as curves resulting
from the intersection of surfaces.

To illustrate this point,
 we display sample cases of such surface plots for the $m_j=\pm 1/2$ and $m_j=\pm 3/2$
states in Figs.~\ref{figa} and~\ref{figb}, respectively. The
intensity of both lasers is  taken to be same, so that the energy
level shift is proportional to the sum of two polarizabilities. This
is plotted on the z axis.  The trap and control laser wavelengths
are given on the x and y axes, respectively. The total
polarizability of the $5p_{3/2}$ state depends
 upon its $m_j$ quantum number, and it is calculated as a sum or difference of the
scalar $\alpha_0$ and tensor $\alpha_2$ polarizabilities, i.e.
$\alpha(5p_{3/2})=\alpha_0-\alpha_2$ for $m_j=\pm 1/2$ states
 and $\alpha(5p_{3/2})=\alpha_0+\alpha_2$ for $m_j=\pm 3/2$ states.
Therefore, we discuss the results for $m_j=\pm 1/2$ and $m_j=\pm
3/2$ states separately. We also find some appropriate trap and
control laser wavelength combinations that have  similar magic
wavelengths for each $m_j$ state. As illustrated by Figs.~\ref{figa}
and ~\ref{figb}, there is a large number of possible combinations of
trap and control wavelengths that will result in the same ac Stark
shift of both levels.

In Table~\ref{tab1}, we list  a number of sample trap and control
wavelength combinations which can be used for state-insensitive
trapping of Rb atoms in $5s$ and $5p_{3/2}$ states. Out of a number
of combinations found, we list only those where one of the laser
wavelengths is twice the other.  This is a case of particular practical interest, since it is attainable by
frequency doubling of the
longer-wavelength laser. The combinations are listed for various
trap and control laser intensity ratios as indicated
 to illustrate the ability to tune the magic wavelength pairs by varying
 the relative intensities.

\begin{figure}[htb!]
  \includegraphics[width=4.5in]{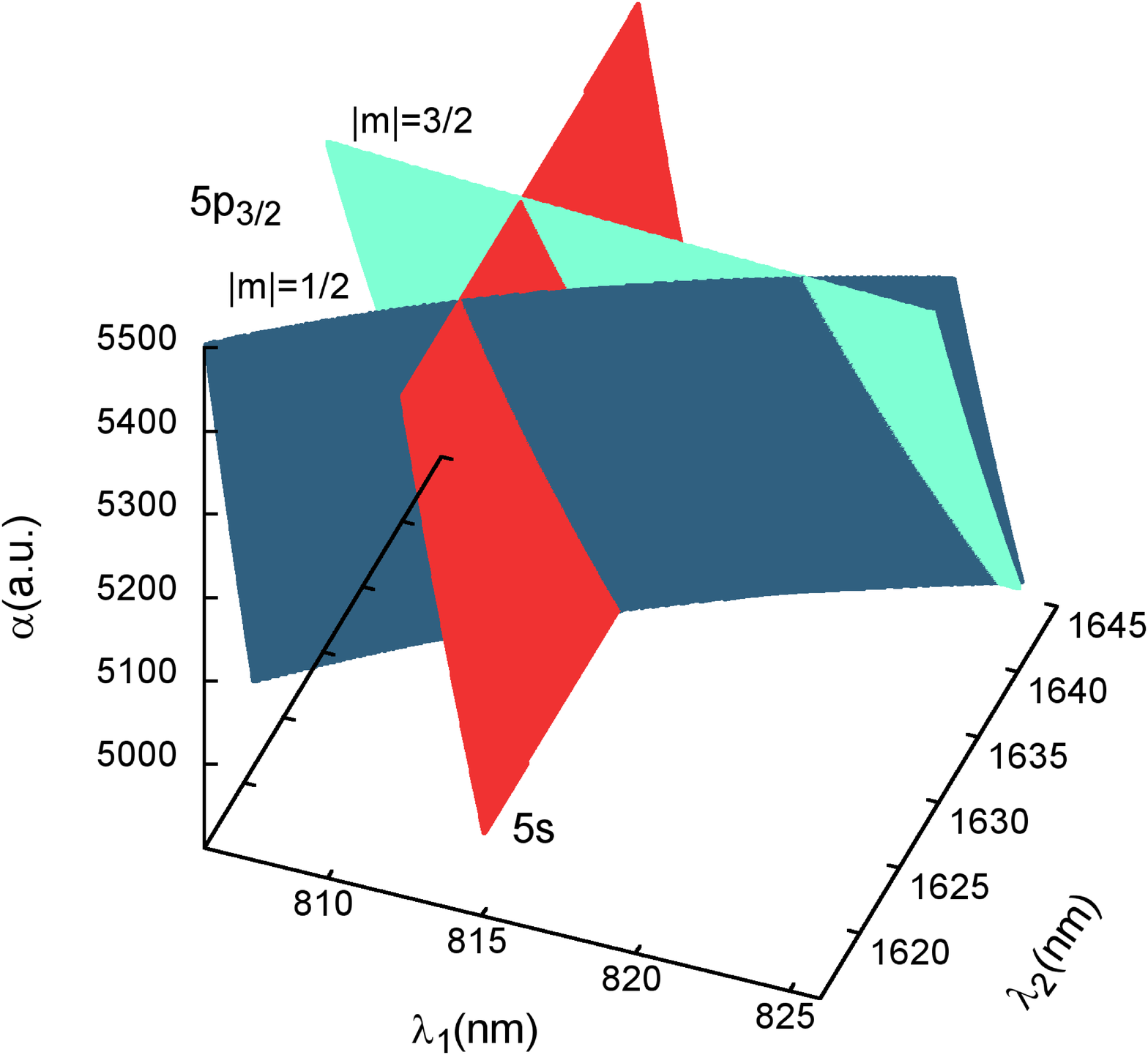}
  \caption{Magic wavelength
pairs for $\lambda_{\rm{1}}=800-830$~nm and $\lambda_{\rm{2}}=1600 - 1660$~nm and equal intensities of both lasers.}
 \label{fig}
\end{figure}

The percentage difference between total polarizabilities of the $5s$ and $5p_{3/2}$ states for the cases listed in
Table~\ref{tab1} is less than 1~\% taking into account uncertainties

 We discuss the magic wavelengths for the $m_j=\pm 1/2$ states
first. For equal  intensities, the combination with $\lambda_{\rm{1}}=788$~nm and $\lambda_{\rm{2}}=1576$~nm, may be
particularly useful since the resulting polarizability is positive and the atoms in red detuned traps are attracted
towards the maximum of the field intensity~\cite{markus,puppe}. Applying a control laser with double the trap laser
wavelength creates a deeper trapping potential for the atom in the ground state and minimizes the difference between
the Stark shifts for ground and excited states. For the $5p_{3/2}, m_j=1/2$ state the polarizability is negative at
788nm ($-10279~a^3_0$) and larger and positive at 1576~nm (15194~$a^3_0$). The uncertainties in the polarizabilities at
combinations which are close to resonance wavelengths are generally higher. The value of $\alpha^{\rm{sum}}$ for some
of the combinations in Table~\ref{tab1} is negative, so that the atoms become low-field seekers. A number of groups
have suggested blue detuned or dark optical traps where
 atoms are surrounded by repulsive light fields and; therefore,
 are captured in dark regions without light ~\cite{puppe,blue1}.
In contrast to the monochromatic case, where very few convenient magic wavelengths were found for $m_j=\pm 3/2$ states,
a number of ``dark'' magic wavelengths for Rb are found in the present bichromatic treatment.

\begin{figure}[htb!]
  \includegraphics[width=3.8in]{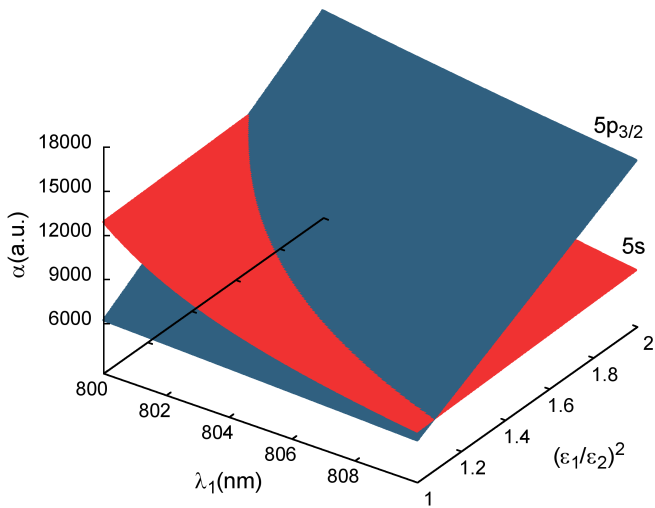}
  \caption{Magic wavelength for the $5s$ and
$5p_{3/2}\mbox{ }m_j = \pm 1/2$ states for $\lambda_{\rm{1}}=800-810$~nm and $\lambda_{\rm{2}}=2\lambda_{\rm{1}}$  for
various intensities of both lasers. The intensity ratio $(\epsilon_1/\epsilon_2)^2$ ranges from 1 to 2. }
 \label{fig4}
\end{figure}

 We
also found a few laser wavelength combinations that support state-insensitive simultaneous trapping for all $m_j$
states. Examples of such cases are grouped together  in a first few rows of Table~\ref{tab1}.  The magic wavelength
combination  for $|m_j| = 1/2$ case is given first, and the corresponding $|m_j| = 3/2$ magic wavelength combination is
given in the following row. We illustrate the example of such magic wavelength combinations (listed in rows 3 and 4 of
Table~\ref{tab1}) in Fig.~\ref{fig},
 where we
 plot
surfaces of the $5s$ and $5p_{3/2}\mbox{ }|m_j| = 1/2, 3/2$ state polarizabilities for $\lambda_{\rm{1}}=806-826$~nm
and $\lambda_{\rm{2}}=1615 - 1645$~nm. The intensities of both lasers are taken to be equal. The magic wavelength by
may be further tuned by adjusting intensity ratio of the two lasers as illustrated in Fig.~\ref{fig4}. The magic
wavelength for the $5s$ and $5p_{3/2}\mbox{ }|m_j| = 1/2$ states for  $\lambda_{\rm{1}}=800-810$~nm and
$\lambda_{\rm{2}}=2\lambda_{\rm{1}}$  are shown for various intensities of both lasers. The intensity ratio
$(\epsilon_1/\epsilon_2)^2$ ranges from 1 to 2.

\section{Conclusion}
In summary, we have explored a bichromatic scheme for state-insensitive optical trapping of Rb atom. Due to the
extensive development of first principles atomic structure theory, semiempirical corrections,  and computational
methodology we are able to explore a wide range of parameter space with reasonable confidence in the uncertainties of
our calculations. We have recently completed a comprehensive survey of calculations of DC polarizabilities,
 for which there exist copious experimental data for comparison within clearly-stated ranges of uncertainty \cite{MSC}.
 In this paper, we specifically explored a case of the Rb atom, where the magic wavelengths associated
with monochromatic trapping were sparse and relatively inconvenient. We have found that the bichromatic approach yields
a number of promising wavelength pairs which are discovered with straightforward parameter choices such as equal laser
intensities  and $\lambda_{\rm{2}}=2\lambda_{\rm{1}}$. The methodology developed in this work allows us to explore
specific cases of interest that may arise in the future experiments where it is essential to precisely localize and
control neutral atoms with minimum decoherence.

This research was performed under the sponsorship of the US
Department of Commerce, National Institute of Standards and
Technology, and was supported by the National Science Foundation
under Physics Frontiers Center Grant PHY-0822671.

\end{document}